# X-ray Emission Spectrum of Liquid Ethanol – Origin of Split Peaks

Osamu Takahashi[1], Mathias P. Ljungberg[2], Lars G. M. Pettersson[3]

[1] Institute for Sustainable Sciences and Development, Hiroshima University, 1-3-1, Kagamiyama, Higashi-Hiroshima, 739-8526 Japan, shu@hiroshima-u.ac.jp

[2] Donostia International Physics Center, Paseo Manuel de Lardizabal, 4, E-20018 Donostia-San Sebastian, Spain

[3] Department of Physics, AlbaNova University Center, Stockholm University, S-106 91 Stockholm, Sweden

**Abstract**

The X-ray emission spectrum of liquid ethanol was calculated using density functional theory and a semi-classical approximation to the Kramers-Heisenberg formula including core-hole-induced dynamics. Our spectrum agrees well with the experimental spectrum. We found that the intensity ratio between the two peaks at 526 and 527 eV assigned as 10a' and 3a'' depends not only on the hydrogen bonding network around the target molecule, but also on the intramolecular conformation. This effect is absent in liquid methanol and demonstrates the high sensitivity of X-ray emission to molecular structure. The dependence of spectral features on hydrogen-bonding as well as on dynamical effects following core-excitation are also discussed.

**Introduction**

X-ray emission spectroscopy (XES) is a powerful technique to obtain atom-specific information on the occupied states in a material[1,2]. In the experiment one detects the photons emitted as a valence electron decays to fill a core-hole created in an x-ray absorption process. For a 1$s$ core hole this results in a direct measurement of the 2$p$ character in the valence level and the energy difference between the core-level and the valence-level. The process is thus inherently photon-in/photon-out and is most appropriately described in terms of an inelastic scattering process. The final state corresponds to a valence ionization or excitation, but where the involvement of the core-level results in a projection on the atom on which the core-hole was created. However, the finite core-hole life-time can lead to dynamical effects[3-7] or life-time vibrational interference[8,9] as different intermediate states can contribute. These effects will be particularly important for hydrogen-bonded (H-bonded) liquids where the light mass of the proton and attraction to the accepting species can lead to particularly large and important effects[10].

The XES spectrum of both liquid water[2,6,11-18] and simple alcohols[19] show a split in the peak at highest emission energy where the relative intensities of the two peaks is affected by isotope substitution which is a clear indication of nuclear dynamics or vibrational interference effects induced by the creation of the O1$s$ core-hole. The interpretation of the split in terms of either final state effects, *i.e.* molecular fragments and intact molecules, or initial state effects, *i.e.* differently H-bonded structures, has caused debate[20,21]. It is important to settle this debate and from a theoretical perspective the simple alcohols provide an attractive test case since only one OH-group is involved in H-bonding.

We have recently reported a study of XES on liquid methanol[22] where we found that the intensity in the two split peaks could be understood in terms of the initial H-bonding structure and the core-hole-induced dynamics which was found to be determined by the initial H-bonding situation. The experimentally observed differences between normal methanol and the deuterated counterpart were excellently reproduced, but the dynamics was not found to give any new peaks in the spectrum. Here we extend these studies to liquid ethanol which confirm the conclusions from methanol with additional complexity due to the possibility of having different conformers in the liquid.

**Methods**

The computational scheme was the same as that applied for methanol by Ljungberg *et al.* [22]. In short, clusters containing 17 ethanol molecules were extracted from a classical molecular dynamics (MD) simulation at room temperature[23]. The oxygen atom of the central molecule is core-ionized. The simulation box contained 1000 ethanol molecules and our sampling was performed on one snapshot, named "snapshot1", by extracting cluster structures around 200 randomly selected molecules. To check the validity of our sampling, the number of H-bonds involving the central molecule was examined. Depending on the number of H-bonds, the sampled clusters were classified as *DmAn*, where *Dm* and *An* correspond, respectively, to the number of donated and accepted H-bonds of the central molecule. To compare our sampling with the statistics in the simulation the number of H-bonds that each of the 1000 molecules was involved in was determined for five disconnected snapshots. The H-bond definition was that proposed by Wernet *et al.*[24] where an H-bond is considered to exist if $r_{OO} < r_{max} - 0.00044\theta^2$ where $r_{OO}$ is the O-O distance, $\theta$ is the angle (radians) between the internal OH and O-O direction and $r_{max}$ is taken as 2.9 Å. The results are summarized in Figure 1. The H-bond distribution in our 200 randomly selected structures is clearly comparable to that of the full snapshots. Thus our sampling is sufficient statistically.

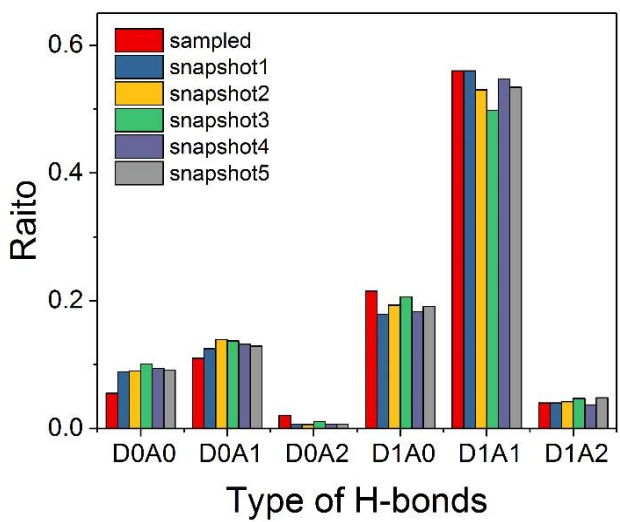

**Figure 1:** Comparing H-bond statistics for the 200 randomly selected ethanol molecules (red) with those of the full simulation box for several snapshots. See text for definition of the nomenclature.

X-ray emission spectra can be computed by the Kramers-Heisenberg (KH) formula[1], which is a one-step description of the absorption of an incoming photon, creating a core-excited state, and the emission of an outgoing photon where an electron falls down to occupy the core orbital. In some cases, these two processes can be separated[25], but this neglects interference between intermediate states. Such interference effects turn out to be crucial in order to describe core-hole-induced vibrational effects[8], which are especially strong in the case where the intermediate potential energy surface is strongly modified compared to the ground state potential. The procedure to handle these effects is the same as applied for methanol[4,22] and will only be briefly summarized here.

The creation of the core-hole projects the ground state OH vibrational wave function onto the core-ionized potential energy surface in a Franck-Condon picture. Subsequently the resulting wave packet starts to evolve on this surface where it eventually decays to the final valence-hole state with the emission of an x-ray photon. The multi-dimensional wave packet evolution is modeled through classical dynamics of the system with the core hole with initial conditions from sampling two positions in the OH quantum mechanical vibrational probability distribution and two momenta in its Fourier transform where, for each, both positive and negative directions were included; the total number of trajectories for each structure was thus eight where each was run for 40 fs in steps of 0.25 fs. Spectra were calculated at each geometry along each trajectory and a total spectrum including the important effects of vibrational interference was generated according to the semi-classical Kramers-Heisenberg (SCKH) approximation to non-resonant XES developed previously[4]. The final calculated spectra are shifted by -2.0 eV to match the experimental spectrum. This error is due to the functional[26] used as well as neglect of relaxation in the valence-ionized final states through the use of orbital energies to estimate their energy position[22].

All spectra are area normalized within the shown range.

Basis sets and density functional were the same as used earlier for methanol[22]. The calculations were performed using the deMon2k density functional program[27].

**Results and discussion**

Figure 2 compares the experimental nonresonant XES spectrum from Schreck *et al.*[19] with spectra computed with (black) and without (dashed) taking into account life-time vibrational interference or core-hole-induced dynamics. It is immediately clear that including dynamics is essential for obtaining the correct intensity relation between the peaks in the spectrum, but the number of peaks and the peak positions are qualitatively in agreement with experiment already in the initial configurations (t=0) of the 200 sampled structures. Including the dynamics due to the creation of the core hole leads to a dramatic change and improvement of the computed spectrum in comparison with the experiment. In particular, the ratio between the intensities of the two peaks at highest emission energy, 3a" and 10a', is reversed with the peak at the 10a' position becoming the strongest. Compared to experiment the computed spectrum shows too little intensity in the 3a" peak. All in all, the spectrum computed with sampling the multidimensional wave packet propagation using classical trajectories with quantum initial conditions[4] shows excellent agreement with the experiment as previously found for liquid methanol[22]. We note that the overall spectrum is somewhat compressed compared to experiment which is due to the neglect of valence hole relaxation effects as the valence binding energies are estimated from the Kohn-Sham orbital energies; these relaxation effects become increasingly important when going from outer valence to the deeper lying orbitals.

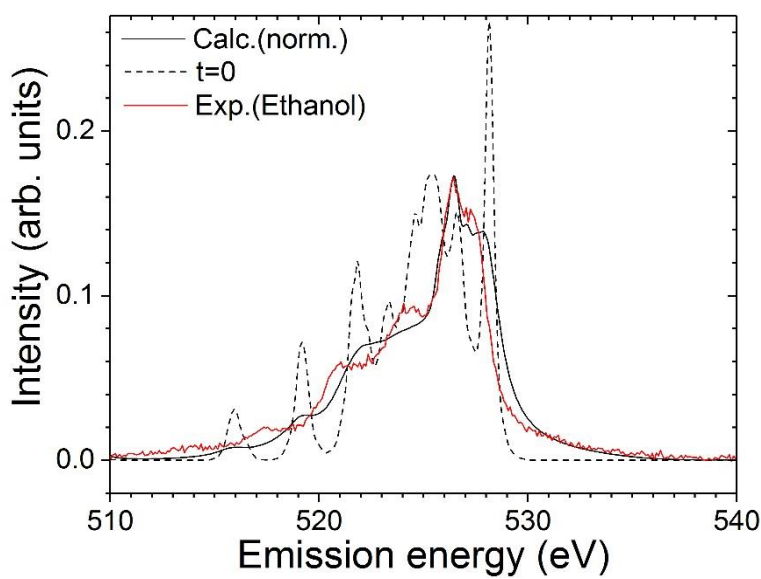

**Figure 2:** Comparison of experimental[19] (red) and computed (black) non-resonantly excited XES for liquid ethanol. The spectrum computed without dynamics, *i.e.* at the initial geometry (t = 0) is indicated with dashed lines.

In Figure 3 we show the computed spectrum decomposed in orbital contributions where the 17 highest orbitals are assigned as 3a", the 17 next as 10a' and so on. The sum of the contributions at each emission energy gives the total spectrum contribution. The inclusion of dynamics broadens the lower-lying states symmetrically, but for the 3a" state there is a significant broadening towards lower energy similar to what has been found earlier for both methanol[22] and water dimer[9]. This forms an underlying background to the 10a' peak, which enhances this feature. Overall, the dynamics broaden and smear out sharp features except for the highest states but we do not observe new features due to dissociation although in some of the trajectories this occurs.

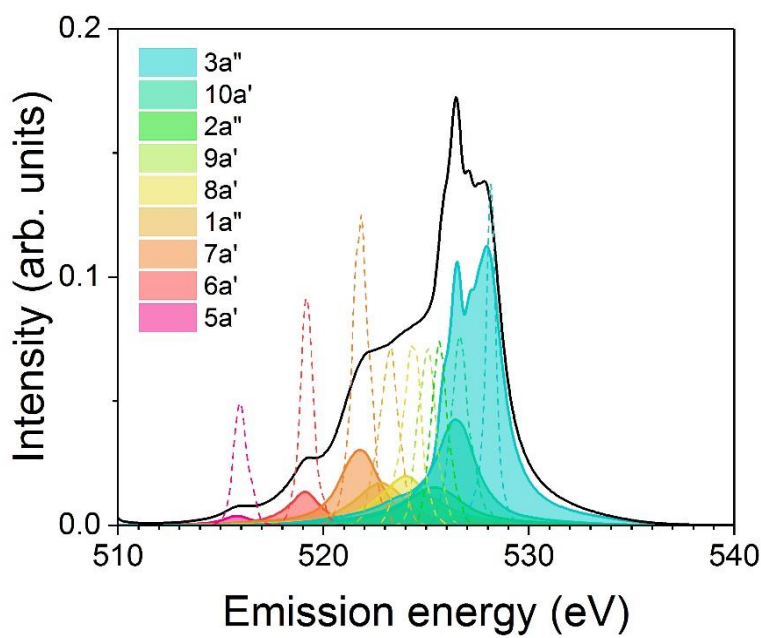

**Figure 3:** Orbital contributions including core-hole-induced dynamics. In particular, the 3a" orbital is asymmetrically broadened towards lower emission energies giving together with the 10a' orbital the enhanced intensity around 527 eV. The computed spectrum (black) is the sum of the orbital contributions at each energy. Dashed peaks give positions without dynamics (t=0).

In figure 4 we investigate the effects of the initial quantum conditions where two internal OH distances are combined with the four initial conditions for the proton momentum. A negative momentum indicates motion towards the OH oxygen and positive away from it;

since the proton is initially bound, both positive and negative momenta are sampled to correspond to the zero expectation value of the momentum. We note first insignificant effects on the initial spectra (t=0) from the variation in internal OH distance (long and short meaning the equilibrium distance plus or minus 0.043 Å). A negative initial momentum gives initial motion towards the oxygen and thus less elongation of the internal OH during the core-hole life-time. The ratio between the 10a' and 3a'' peaks is clearly determined by the dynamics where, independent of initial OH distance, the 10 a'' peak is enhanced by increasingly positive dynamics. For the long initial OH distance the negative initial momenta lead to the decay occurring largely within the molecular region and thus less effects on the 3a'' state which remains rather sharp. In conclusion, for the sampling of initial conditions the momentum distribution is the more important as earlier found for simulations of XES spectra of water[4].

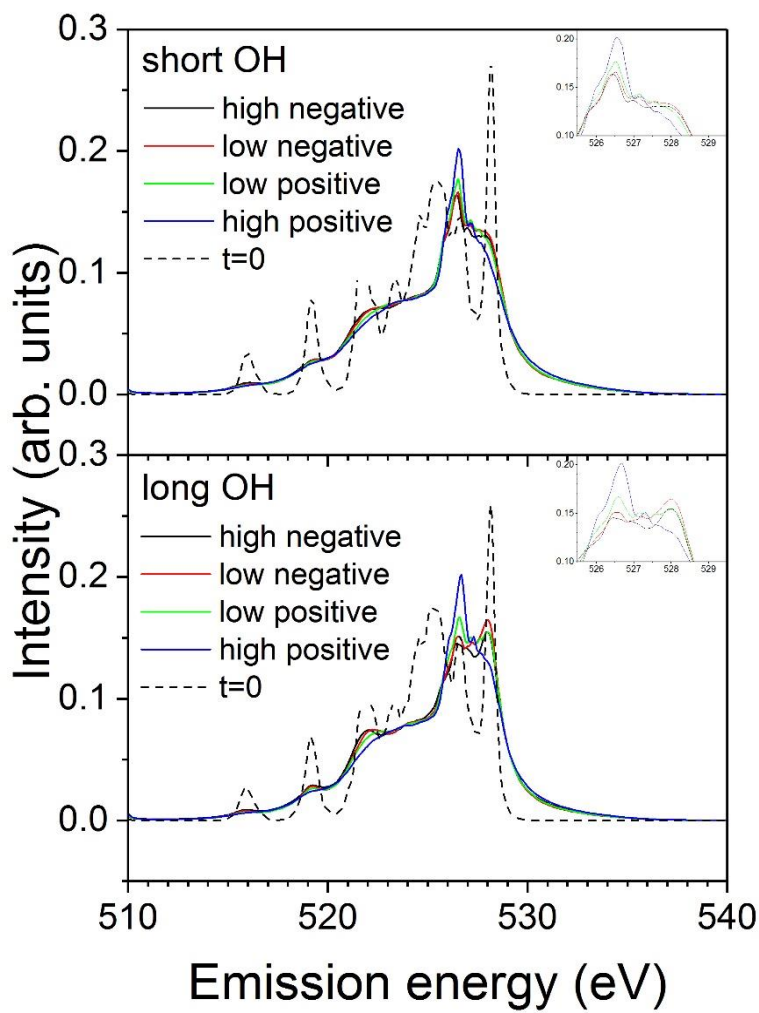

**Figure 4:** Effect of initial conditions. (Top) Sampling of initial momenta in the case of initial internal OH distance shorter than the equilibrium. Negative momentum has the hydrogen moving towards its oxygen and positive momentum moving away from it. (Bottom) Same as above, but for initial sampled OH distance longer than equilibrium. The inserts show the peak contributions between 526 and 530 eV. Dashed spectrum is without dynamics (t=0).

We now turn to the sampling of H-bond situations in the simulated liquid, where in figure 5 the sensitivity to H-bond length for acceptors and donors is analyzed and in figure 6 the corresponding sensitivity to H-bond angle. From figure 5 it is clear that the initial structures (t=0) give very similar spectra if one doesn't take dynamics into account. However, the initial structure determines the dynamics, as is most clearly seen for the H-bond donors where short H-bond distances enhance the 10a' peak significantly. This is similar to the case of water dimer[9] where the presence of the accepting oxygen leads to an attractive potential for proton motion between the two oxygens after core-ionization of the donor. The proton moves back and forth in this potential, which results in emission over the entire H-bond distance and a strong, asymmetric broadening of the 3a" state towards lower emission energy, thus enhancing the 10a' peak. For long H-bond distances the spectrum remains more similar to the initial structure since in this case there is no attractive potential for the proton which either remains localized or dissociation occurs for the OH group where the O-C bond breaks[22]. Acceptors show similar effects albeit with stronger differences between the different classes where again long H-bonds generate significantly less dynamics. From figure 6 it is clear that the more well-defined H-bonds in terms also of the angle induce more dynamics as evidenced by comparison with the spectra at t=0. The effects of the angle are most pronounced for the H-bond donors, with the 10a' peak increasing with decreasing deviation from a straight H-bond. In conclusion, we thus find that the dynamics is strongly dependent on the initial structure in terms of H-bonding which makes XES combined with simulations that take both structure and dynamics into account a sensitive probe of the local structure in the liquid.

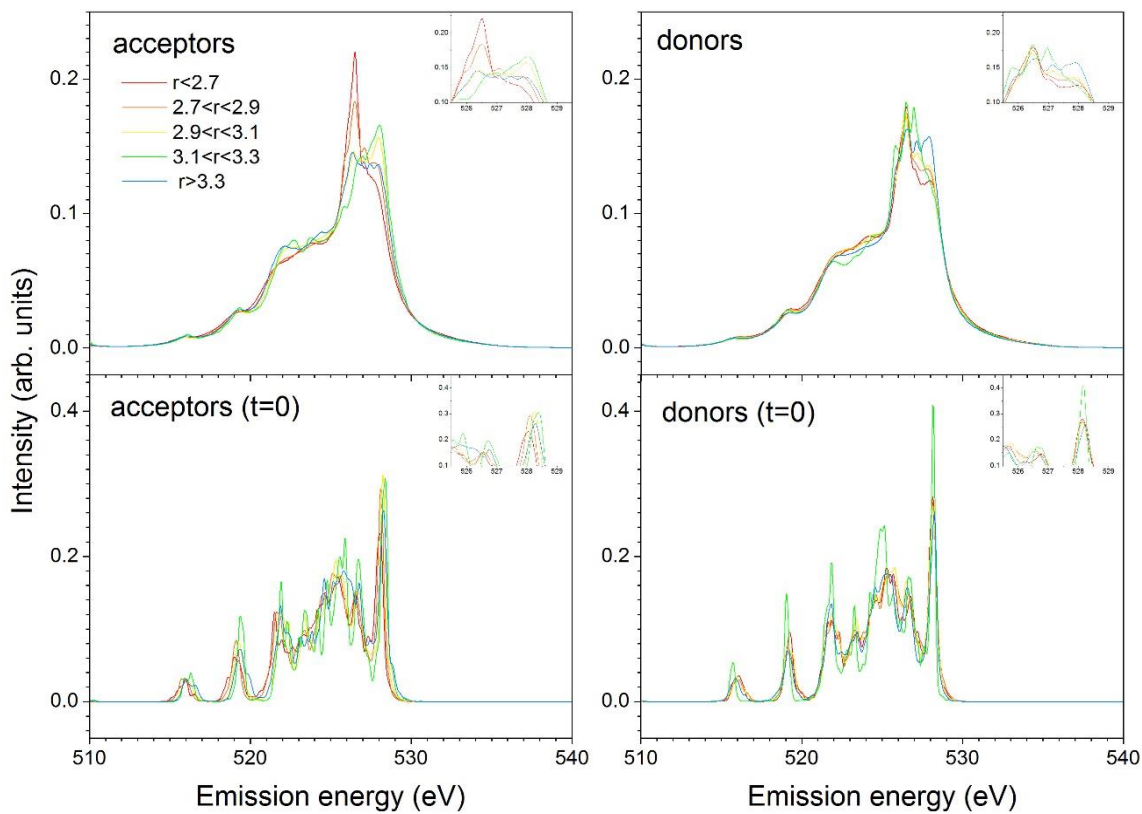

**Figure 5:** Dependence of computed XES on initial structure. (Left panels) Accepted H-bond distance with (Top) and without (Bottom) dynamics. (Right panels) Donated H-bond distance with (Top) and without (Bottom) dynamics. Inserts show peak structure in the energy range 526 to 530 eV.

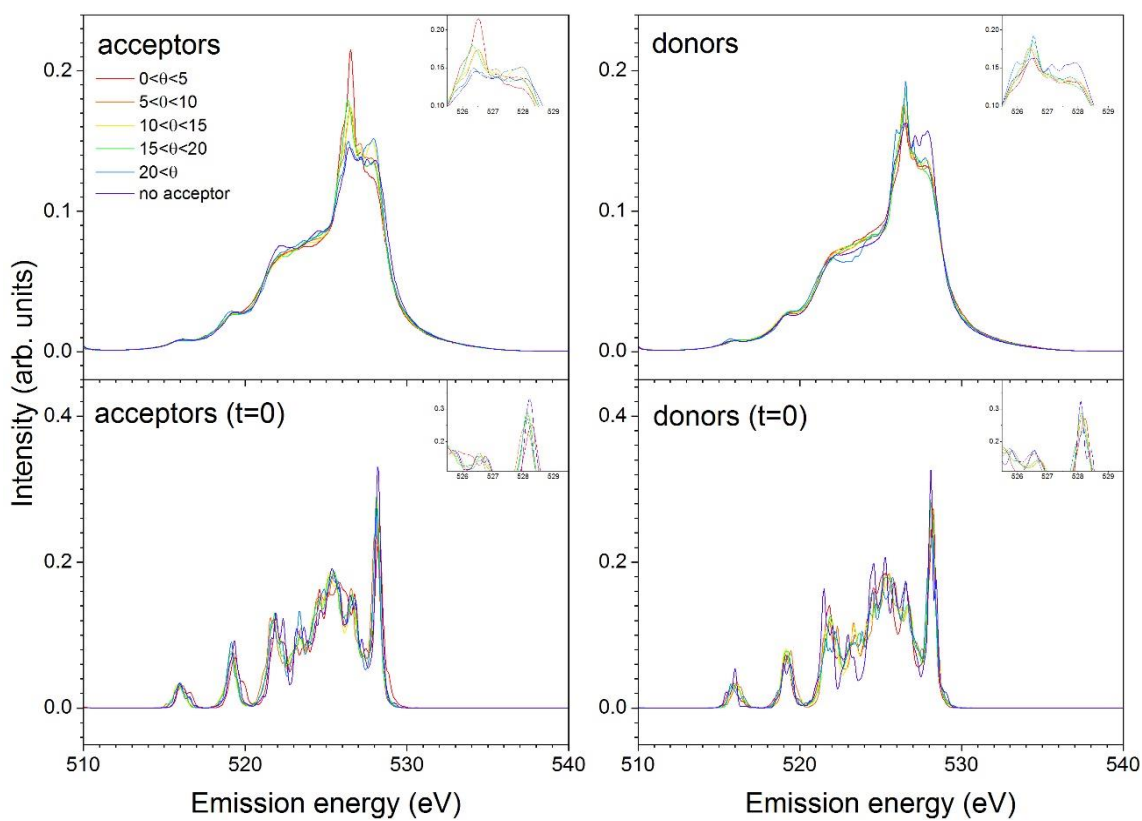

**Figure 6:** Dependence of computed XES on initial structure. (Top) Accepted H-bond angle and (Bottom) donated H-bond angle. Inserts show peak structure in the energy range 526 to 530 eV.

For ethanol two structural isomers, *anti* and *gauche* (see figure 7), exist which dominate in the simulated liquid with less probability to find structures in between. In gas phase the energy difference between them has been estimated as only 40 cm$^{-1}$ favoring the *anti* conformer[28-30], but the double degeneracy of the *gauche* conformer still leads to a gas phase ratio of 42:58 favoring *gauche*; an analysis of the origin of the energy difference has been given by Scheiner and Seybold[31].

In figure 8 we observe clear differences in the spectra of the two isomers, particularly when including dynamics. The *anti* conformer gives strongly enhanced 10a' peak and depleted 3a'' while for the *gauche* conformer the 3a'' peak remains more well-defined and, in addition, a peak around 524 eV becomes visible which is only somewhat broadened by dynamics. Combined with the lack of intensity from the *anti* conformer in this region this peak is lost in the summed spectrum, However, based on comparison between the ratio between 3a'' and 10a' in experiment and calculation we can speculate that in the MD simulation the fraction of *gauche* conformers is slightly underestimated.

**Conclusions**

In the present paper, we have investigated the XES spectrum of liquid ethanol using a semi-classical approximation to the Kramers-Heisenberg formula within the framework of DFT with the core-hole-induced dynamics included through classical trajectories sampling quantum initial conditions in the O-H zero-point distance and momentum distributions. Our calculated spectra describe well the experimental spectrum. The effects on the spectral features due to the dynamical effects following core-excitation are examined in terms of initial momentum distribution, hydrogen-bonding structure, and valence orbital decomposition. In agreement with our earlier study on liquid methanol[22], several effects affect the spectral shape of XES; (i) a static core-level shift, (ii) the sampling of quantum initial conditions, and (iii) dynamical effects due to the core-hole. In contrast to the case of methanol, we find that for ethanol (iv) also the intramolecular conformation affects the spectral shape. (i) and (ii) are directly related to the initial structure, while (iii) and (iv) depend on the dynamical behavior following the core-excitation in the initial structure.

Although the correspondence between the present theoretical spectra and the experiment is quite good, there are still some improvements that may be considered. For the lower emission-energy region, the theoretical spectrum is compressed compared to experiment. The origin of this compression is the lack of relaxation effects which increase for the deeper valence orbitals, since the analysis is based on orbital energies for the valence final hole states. To include this effect, core-level spectroscopy calculations by time-dependent DFT (TDDFT) or the use of newer exchange-functionals may be considered[32,33], although at significantly increased cost considering the sampling in terms of core-hole-induced trajectories with quantum initial conditions that must be performed. We note that only the proton dynamics related to H-bonding was considered with quantum initial conditions in the present study while all other degrees of freedom were treated classically. The good agreement with experiment we find within this restriction indicates that dynamical effects for other degrees of freedom in the molecule itself are less important.

We conclude that the dynamical effect which can be included through the use of the SCKH method with correct statistical geometry and momentum sampling is needed to correctly reproduce the features of the XES spectrum of H-bonded liquids. Our procedure can be applied, not only to small molecular systems, but also to the multidimensional condensed phase represented by liquids with a connecting H-bonded network.

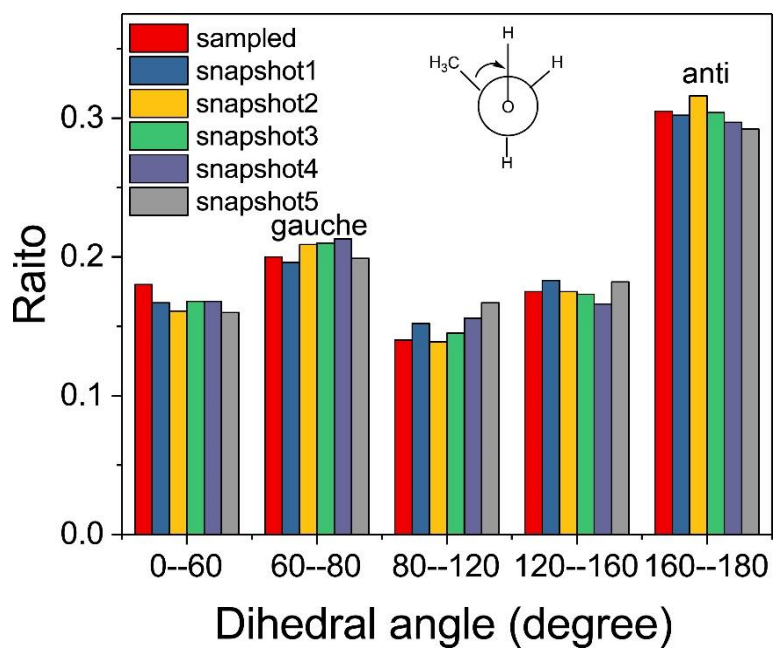

**Figure 7:** Definition and distribution of *gauche* and *anti* conformers.

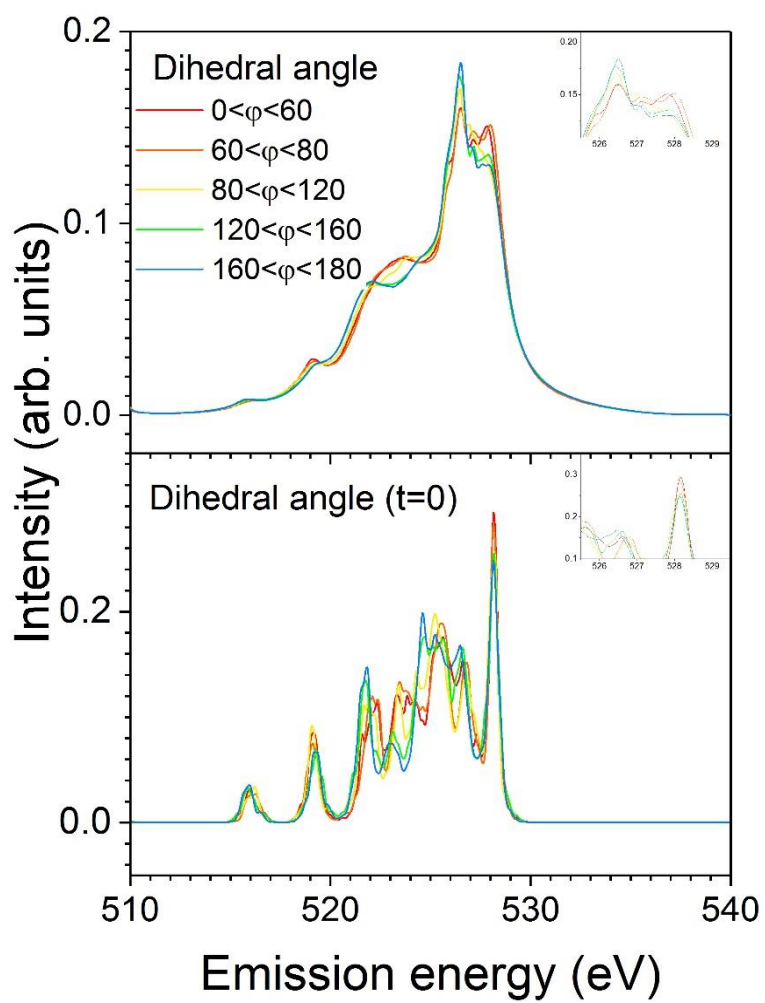

**Figure 8:** Dependence on dihedral angle.

# Acknowledgements

We thank Mikko O. Hakala and Susi Lehtola for providing structures from their molecular dynamics simulations of liquid methanol, and Simon Schreck for sharing the experimental spectra. O.T. thanks Prof. Takayuki. Ebata for helpful support. The authors would also like to acknowledge the Information Media Center at Hiroshima University for the use of a grid with high-performance PCs and the Research Center for Computational Science, Okazaki, Japan for allowing the use of Fujitsu PRIMERGY. This work was supported by JSPS KAKENHI (No.15K04755).